\documentclass[preprint]{aastex}
\usepackage{natbib}
\shortauthors{Gizis}
\shorttitle{AKARI M Dwarfs}

\begin{document}

\title{M Dwarfs in the Mid-Infrared AKARI/IRC Sky Survey}

\author{John E. Gizis}
\affil{Department of Physics and Astronomy, University of Delaware, 
Newark, DE 19716}
 
\begin{abstract}
We present a sample of 968 K5-M6 dwarfs detected in the AKARI-IRC Point Source Catalog. The vast majority of these low-mass stars have brightnesses in the AKARI broad 9-micron (S9W) filter that match expectations of model photospheres.  Mismatched double stars have been excluded. Hot  (200-600K) debris disks can produce excesses in this mid-infrared filter.  We discuss five M dwarfs which have excesses that may be significant but require independent confirmation. We also discuss 50 detections with the L18W filter, and support the claimed Spitzer detection of a warm debris disk around AT Mic. We also report a previously unrecognized disk around a T Tauri star.  
\end{abstract}


\section{Introduction}

Since the IRAS sky survey, mid-infrared astronomy has revealed both protoplanetary circumstellar disks around young stars \citep{1987ARA&A..25..521B} and "debris disks" around older stars \citep{2008ARA&A..46..339W}.  The most common type of star, representing masses from $\sim 0.08$ to $\sim 0.6 M_\odot$, is the M dwarf \citep{2005-book-reid}. Although the tremendous luminosity of A stars helps make their disks relatively easily observable, and G dwarfs are similar to our own Sun, a complete understanding of stars and their planetary systems requires an accounting of M dwarfs. 

Despite their frequency --- some two thousand M dwarfs within 25 parsecs are known \citep{1995AJ....110.1838R} --- their low luminosities ($L < 0.1 L_\odot$ ) make them difficult mid-infrared targets. 
In their re-analysis of IRAS data, \citet{2007ApJ...660.1556R} report that only one M dwarf, AU Mic (Gl 803), was detected.  AU Mic has a debris disk that has also been detected in the submillimeter \citep{2004ApJ...608..526L} and imaged by Hubble Space Telescope \citep{2004Sci...303.1990K}. Earlier IRAS-based studies that claimed numerous detections and a systematic difference between active (dMe) and inactive stars (dM) are apparently incorrect. \citet{2002AJ....124..514S}, for example. showed apparent excesses in five M dwarfs were false. \citet{2005ApJ...631.1161P} also found no excess at 11.7 microns with the Keck Telescope in a sample of nine M dwarfs that were either young or reportedly had IRAS excesses. The Spitzer Space Telescope has been used to study M dwarfs at high precision, but the pointed telescope has been limited to dozens of M dwarfs.  \citet{2006ApJ...650.1133R} found that eight M dwarfs show no evidence of an infrared excess due to chromospheric activity, while \citet{2007ApJ...662.1245M}'s  mid-infrared spectra of 10 M dwarfs showed no expected features or emission lines. \citet{2007ApJ...667..527G} found no excesses at 24 microns in a sample of 62 M dwarfs.   \citet{2009ApJ...698.1068P}, however, found excesses for the M4.5 dwarf AT Mic and the K7 dwarf Gl 907.1.  \citet{2008ApJ...687.1107F} report six M dwarf debris disk candidates in the $\sim 40$ Myr old open cluster  NGC 2547.  Submillimeter detections of field M dwarf debris disks include the stars AU Mic, Gl 182 \citep{2004ApJ...608..526L}, Gl 842.2 \citep{2006A&A...460..733L}, and perhaps Gl 526 \citep{2009A&A...506.1455L}.
 
The AKARI, or ASTRO-F, satellite \citep{akari} has scanned more than 90\% of the sky at higher resolution and sensitivity than IRAS using the Infrared Camera \citep{akari-irc}. The resulting AKARI-IRC Point Source Catalog \citep{akari-psc} has a typical 5-sigma level of 50 mJy in the S9W (9 micron) and 120 mJy in the L18W (18 micron) filters, though the sky coverage is not uniform.  \citet{akari-nearby} have correlated the catalog with 2MASS \citep{2mass} and Hipparcos \citep{hipparcos} stars. This analysis, for example, led to the discovery of a bright debris disk around HD 165014 \citep{2010ApJ...714L.152F}.  
 
This high quality, mid-infrared sky survey allows us to investigate the properties of nearby M dwarfs.  We present AKARI detections of  stars in the range K5-M5.  In Section~\ref{sec-data}, we present the AKARI matches.  In Section~\ref{sec-discuss}, the implications for M dwarf photospheres and M dwarf debris disks are discussed.  

\section{AKARI data \label{sec-data}}

The S9W and L18W filters used by the AKARI IRC camera are quite broad. As discussed in the AKARI-IRC Point Source Catalogue Release note \citep{akari-psc-note}, the effective wavelengths and flux calibration of AKARI are given for a constant energy ($\lambda^{-1}$) source. M dwarf photospheres will be bluer, and thus have a different effective wavelength. If $R_i$ is the relative response (in electrons per photon) of the filter:

\begin{equation}
f^{quoted}_{\lambda} ={ {\int_{\lambda_1}^{\lambda_2} \! R_i(\lambda)\lambda f_\lambda \, d\lambda} \over 
 {\int_{\lambda_1}^{\lambda_2} \! \lambda^{-1} R_i(\lambda)\lambda \, d\lambda}  }
\end{equation}

\begin{equation}
\lambda_{eff} ={ {\int_{\lambda_1}^{\lambda_2} \! \lambda R_i(\lambda)\lambda f_\lambda \, d\lambda} \over 
 {\int_{\lambda_1}^{\lambda_2} \! R_i(\lambda)\lambda f_\lambda \, d\lambda}  }
\end{equation}

\begin{figure}
\epsscale{0.7}
\plotone{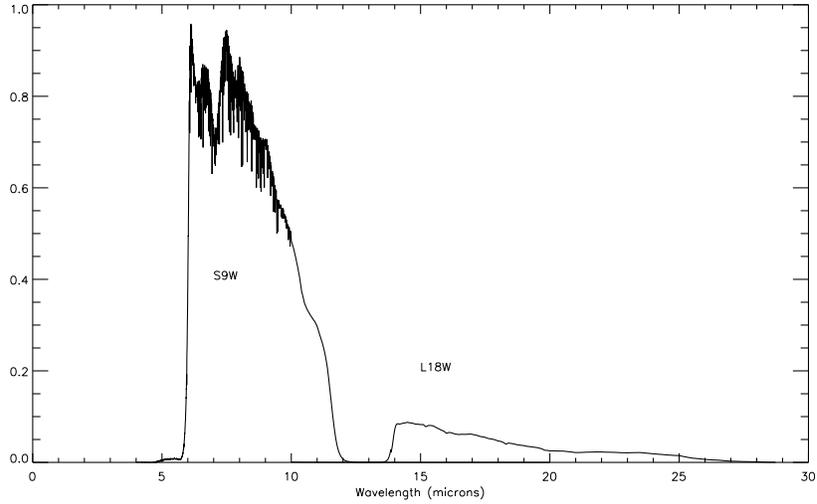}
\caption{The quantity $W=R_i(\lambda)\lambda f_\lambda$ for a theoretical 3600K M dwarf observed through the S9W and L18W filters.  \label{fig-filter}}
\end{figure}

We can quantify this effect using a NextGen \citet{1999ApJ...512..377H}  model of a 3600K solar metallicity star with $\log g=5.0$, corresponding to an M1 or M2 dwarf, and the relative spectra response function ($R_i(\lambda)\lambda$) tables for S9W and L18W  from the AKARI website\footnote{\url{www.ir.isas.jaxa.jp}}.  Figure~\ref{fig-filter} shows the quantity $W=R_i(\lambda)\lambda f_\lambda$. Note the strong contributions from the bluer half of the filter: The effective wavelength for this M dwarf model is just 8.27 microns. As a result, M dwarfs in the AKARI IRC S9W sky survey will be brighter than their actual flux densities at 9 microns.  As also shown in the figure, M dwarfs are considerably fainter at L18W, and furthermore AKARI is less sensitive there, so most known nearby M dwarfs are typically too faint to be detected in L18W. The NextGen model predicts $\lambda_{eff} =18.0$ for L18W. 

It is convenient to work in a magnitude system:

\begin{equation}
m_i = -2.5 \log (F_i / ZP_i)
\end{equation}

where $F_i$ is the flux density reported in the AKARI point source catalog for each filter. We have adopted the zero-points given by \citet{akari-psc}:  56.26 Jy for S9W and 12.00 Jy for L18W. This is a traditional system where A0 stars have zero color. To check these zero-points, I compare the magnitudes of nine A0V and A1V stars selected from \citet{2007PASP..119..994E} in the 2MASS and AKARI surveys (Table~\ref{tab-astars}) that have $m_{S9W} < 6$. The mean $K_s - m_{S9W}$ is -0.005, the median is -0.004, and the standard deviation is 0.05. I conclude the zero-points are appropriate. In this system, a 50 mJy source has $m_{S9W} = 7.63$, so it is evident that only nearby M dwarfs will appear in the AKARI sky survey.  A 120 mJy source has $m_{L18W} = 5.00$. Note that no attempt is made to apply color terms under the philosophy that the reported magnitudes should correspond to observed brightnesses. Models can be used to be interpret the observations. As noted above, the catalog flux densities or magnitudes given for M dwarfs will not correspond to their flux densities at 9.00 microns.   


For a sample of nearby M dwarfs, I primarily use the Palomar/Michigan State Survey (PMSU) survey\citep{1995AJ....110.1838R,1996AJ....112.2799H}, which was based on \citet{pcns3}. The advantage of this sample of over 2000 late-K and M dwarfs within 25 parsecs is that they have uniform spectroscopy. The band-strength index TiO5 is linearly correlated with spectral type over the range K5V-M6V. 

\begin{equation}
Sp = -10.77 \times {\rm TiO5} + 8.2
\label{eqn-sp}
\end{equation}

Values of -1 and -2 correspond to K7 and K5 respectively. Chromospheric activity is measured in the form of H$\alpha$ equivalent width. The nearby star sample at that time is known to have been biased against the inclusion of low proper motion, and hence young or magnetically active, M dwarfs. We supplement our main PMSU sample with a sample of X-ray active dwarfs identified by \citet{2006AJ....132..866R} using 2MASS and ROSAT. These dwarfs have TiO5 and H$\alpha$ on the same system.  

The matching procedure is as follows: The nearest AKARI source within 30 arcseconds is picked. (The large radius is used due to the large proper motion of some stars.) This AKARI source, in turn, is matched to the 2MASS point source catalog. I then examined the 2MASS colors and then the finding charts to determine if the match was correctly made.  A second pass was made to identify stars with $\mu>1$. The 939 K5-M5 dwarfs -- or rather, all M dwarf systems that are unresolved at $\sim 1-2$ arcsecond resolution -- that match AKARI sources are given in Table~2.  The AKARI coordinate-based names have the prefix AKARI-IRC-V1 J.  Although the Table~2 gives spectral types in the form M1, M1.5, M2, etc. as in the cited papers, all following calculations and figures use Equation~\ref{eqn-sp} to continuous spectral types.

One important limitation to the AKARI data is the resolution of $\sim 10$ arcseconds. Although a great improvement over IRAS, many nearby M dwarf binaries are resolved by 2MASS and PMSU spectroscopy but unresolved by AKARI. If just the primary's 2MASS magnitude is compared to the system's AKARI magnitude, the source appears to have a large excess. The mean $K_s - m_{S9W}$ color of single M dwarfs is 0.3. Because M dwarfs are not much redder than background AFGK stars, even an unrelated star that is close on the sky can have a significant effect and cause a false excess.  This occurs more frequently near the Galactic plane. In addition, many PMSU M dwarf companions to brighter primaries are unresolved by AKARI.  All such systems with unresolved, and hence unreliable, AKARI data have been excluded from Table~2.

A number of objects are included in the search sample for historical reasons that are actually more distant, young (T Tauri) stars. The star GJ 2084  (HD 98800, or TWA 4) was included in the PMSU, although it is now known to be a more distant  \citep{hipparcos} member of the young TW Hydrae Association with a circumstellar disk  \citep{2007ApJ...664.1176F}. It matches with AKARI-IRC-V1 J1122052-244639  with $m_{S9W}=4.05$ and $K_s-m_{S9W}=1.27$. It would clearly be identified as have a mid-infrared excess from this data. The remarkable $m_{L18W} = 0.63$  would confirm this. The Riaz et al. sample by its nature included a number of known T Tauri stars with larger excesses ($K_s-m_{S9W}>1.2$), but we have excluded these sources from Table~2.  We consider two such members of the $\eta$ Cha Association \citep{1999ApJ...516L..77M}.  The observed AKARI S9W magnitudes of 6.96 and 5.91 for RECX 1 and RECX 11 respectively are consistent with the Spitzer/IRAC 8 micron magnitudes of 7.11 and 5.97 \citep{2005ApJ...634L.113M}. With L18W, RECX 1 is not detected but RECX 11 has a magnitude of 4.14. This again is consitent with the expectations from Spitzer MIPS \citep{2008ApJ...683..813G}, where RECX11 is magnitude 3.90 at 24 microns. In short, we find that T Tauri type disks are easily detected in the AKARI data.  We leave such objects out of our nearby M dwarf sample. We did note that the K7 \citet{2006AJ....132..866R} source 2MASS J13335481-6536414 /  	1RXS J133355.3-653641 matches AKARI-IRC-V1 1333554-653643, as well as IRAS 13303-6521. There is nearby M0.5 star (2MASS J13335329-6536473) in \citet{2006AJ....132..866R} at 11 arcseconds which may contribute to the AKARI source. With magnitudes $K_s = 7.81$, $m_{S9W} = 5.39$ and $m_{L18W} = 3.74$, even recognizing the nearby M dwarf, 2MASS J13335481-6536414 appears to be a previously unrecognized young T Tauri star with disk.  The lack of significant proper motion supports the T Tauri star hypothesis. We note that the B star HD 117708 is within two arcminutes,but since it is fainter in 2MASS, it is unlikely to be physically associated.   

As a further supplement, we also searched for additional nearby M dwarfs in the AKARI data from the analysis by \citet{2004AJ....128..463R}. Many of these stars lack TiO5 measurements so we simply list them seperately in Tables~\ref{tab-reid}. Twenty-nine stars are listed. Again, confused systems were excluded.  

\section{Analysis\label{sec-discuss}}

\subsection{Photospheres}

Tables~2 and~\ref{tab-reid} list AKARI detections of a total of 950 K5-K7 and M dwarf systems.  Our analysis is primarily based on the 939 systems in Table~2 which have high quality infrared data and spectroscopy. It is  clear that for the vast majority of the target stars the AKARI detection is photospheric. To describe the correlation of spectral type and $K_s - m_{S9W}$ color, we fit a line:

\begin{equation}
K_s - m_{S9W} = 0.262 + 0.056 \times Sp
\label{eqn-fit}
\end{equation}

This agrees well with expectations for M dwarf photospheres. Included in Table~2 is the quantity $\Delta_9$, which is the deviation between the observed color and that expected from Equation~\ref{eqn-fit}. The quantity $N_\sigma$ is $\Delta_9$ devided by the uncertainty reported in the  AKARI catalog. $N_\sigma$ will be overestimated for bright stars with high signal-to-noise since the repeatability of AKARI is said to be 4\%, and since 2MASS has errors of order 2-3\%.  

We also calculated synthetic photometry for NextGen model atmospheres with solar composition and $\log g = 5.0$. Equation~\ref{eqn-temp} is an adequate characterization of the temperature scale for our purposes  \citep{2005-book-reid} .

\begin{equation}
T_{eff} = 3800K - 150 \times Sp
\label{eqn-temp}
\end{equation}

\begin{figure}
\epsscale{0.7}
\plotone{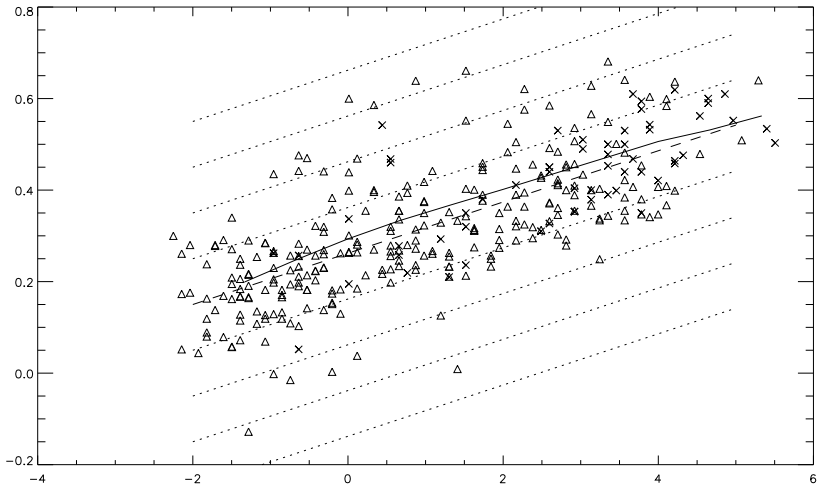}
\caption{Spectral type and $K_s - m_{S9W}$ color.  The triangles are inactive dM dwarfs and the crosses are active dMe dwarfs. On the plot, spectral type -2 is K5, -1 is K7, 0 is M0, 1 is M1, and 4 is M4. The solid line is the predicted color using the NextGen models.  The dashed line is the linear fit to the data, and dotted lines show that fit $\pm 0.1$, $\pm 0.2$, $\pm 0.3$, and $\pm 0.4$ magnitudes . Only stars with AKARI signal-to-noise of 20 or more are plotted.
\label{fig-photo}}
\end{figure}

The predicted relationship is shown in Figure~\ref{fig-photo}. As expected from the small Spitzer samples, the observed mid-infrared spectral energy distribution agrees well with theoretical expectations for M dwarf photospheres. 

\begin{figure}
\epsscale{0.7}
\plotone{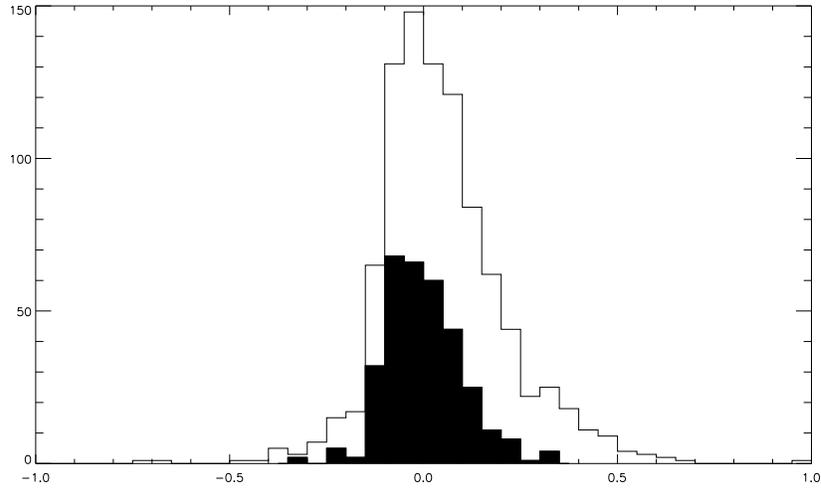}
\caption{A histogram of deviations from the fitted relationship between spectral type and $K_s - m_{S9W}$. The solid shaded histogram counts only stars wth signal-to-noise of 20.0 or greater and the open histogram counts all stars. 
\label{fig-hist}}
\end{figure}

For stars which have an AKARI S9W signal-to-noise of at least 20, the standard deviation about the fitted line is 0.10 magnitudes. If that group is further restricted to stars with $m_{S9W} < 6$, it is only 0.07 magnitudes.  The scatter is plotted in Figre~\ref{fig-hist}. The difference between the observed and predicted (Eqn~\ref{eqn-fit}) color is given in Table~2 as  $\Delta_9$.  $N_\sigma$ in the same table is the divided by the uncertainty in magnitudes due solely to the uncertainties reported in the AKARI catalog. As seen in Figure~\ref{fig-photo}, there is no apparent difference between active (dMe) and inactive (dM) dwarfs.  

\begin{figure}
\epsscale{0.7}
\plotone{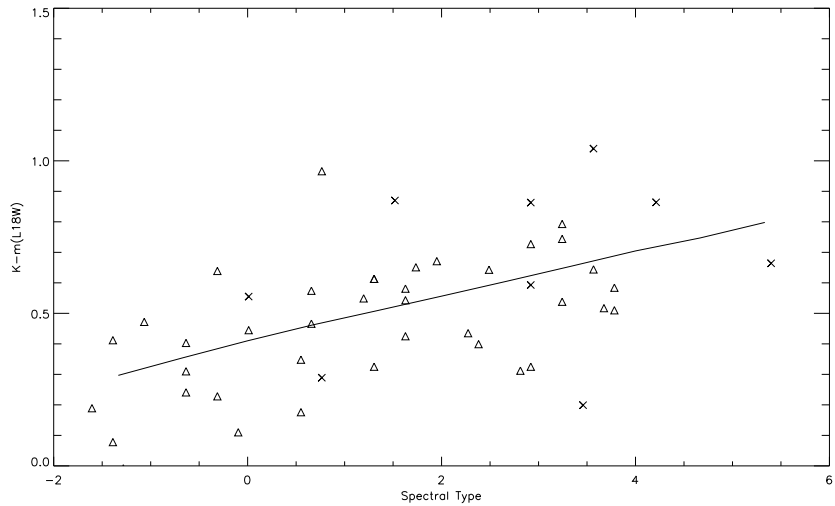}
\caption{Spectral type and $K_s - m_{L18W}$ color.  Symbols as in Figure~\ref{fig-photo}. The solid line is the predicted color using the NextGen models.  Even low signal-to-noise detections are plotted.
\label{fig-photo18}}
\end{figure}

Aside from the T Tauri stars discussed in the previous section, only 50 stars are detected with L18W.  The data are shown in Figure~\ref{fig-photo18} along with the NextGen predictions.   Again, the data match the predictions of model photospheres as expected. This sample necessarily consists of the brightest, closest M dwarfs and therefore overlaps with the pointed Spitzer studies discussed in the introduction. 

\subsection{Excesses and Debris Disks}

\begin{figure}
\epsscale{0.7}
\plotone{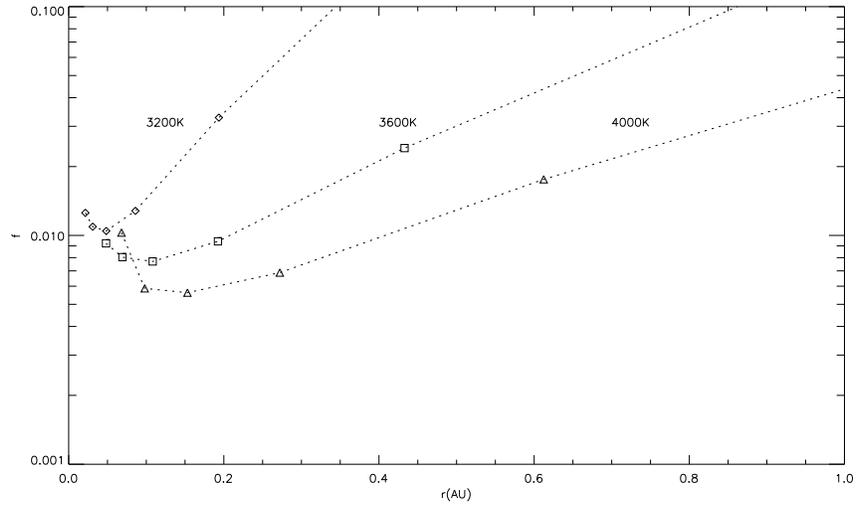}
\caption{The limit on $f$ from a dust disk of temperatures 600K to 100K in steps of 100K assuming $\Delta m_{S9W} = 0.4$. The three stars shown have NextGen spectra of 4000K with $0.1 L_\odot$ (triangles), 3600K with $0.05 L_\odot$ (3600K), with 3200K and $0.01 L_\odot$. In all cases the 100K point lies off the chart with $f>1$.  Disks that lie above the curve could be detected in our sample.  
\label{fig-fdet}}
\end{figure}

\begin{figure}
\epsscale{0.7}
\plotone{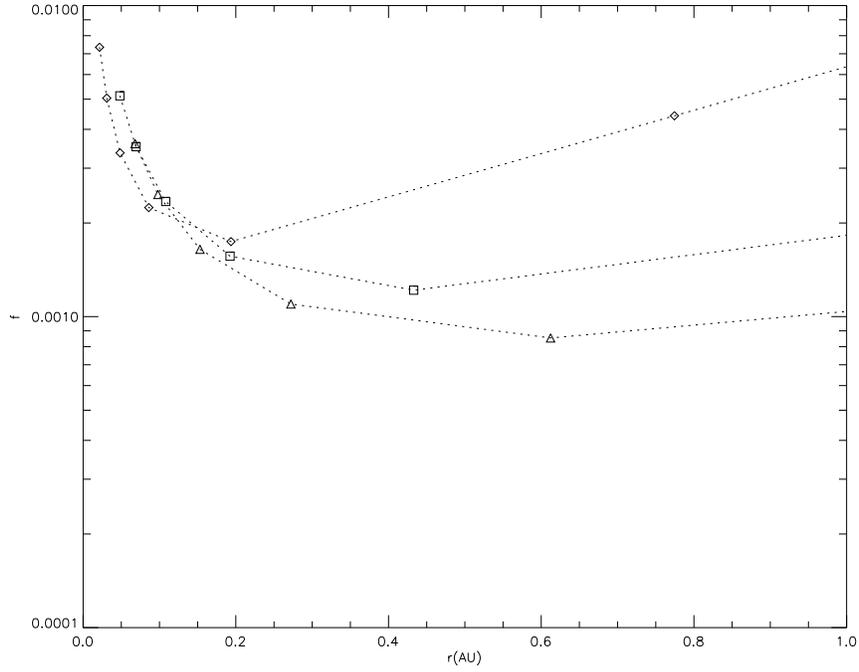}
\caption{The limit on $f$ from a dust disk of temperatures 600K to 100K in steps of 100K assuming $\Delta m_{L18W} = 0.5$. The three stars shown have NextGen spectra of 4000K with $0.1 L_\odot$ (triangles), 3600K with $0.05 L_\odot$ (3600K), with 3200K and $0.01 L_\odot$. An additional point for a disk of 50K lies off the chart to the upper right.  Disks that lie above the curve could be detected in our sample.  
\label{fig-fdet18}}
\end{figure}

Debris disks are reviewed by \citet{2008ARA&A..46..339W}. The AKARI M dwarf data set amounts to a calibration-limited sample that can be used to search for debris disks.  Using Wyatt's equations 3 and 11, we can model our sensitivity limit in the S9W and L18W filters. The simple model assumes the debris disk dust is optically think, at a single temperature, and emits blackbody radiation. However, instead of simply using the ratio ($B_\nu(\lambda,T_\star)[B_\nu(\lambda,T)]^{-1}$) at a single wavelength, we use the ratio of a stellar NextGen model to the blackbody dust as observed through the AKARI filters. The results are shown in Figure~\ref{fig-fdet} for three model stars assuming a detection limit of 0.4 magnitudes above the photosphere at S9W and Figure~\ref{fig-fdet18} for L18W assuming a detection limit of 0.5 magnitudes. $f$ is $L_{IR}/L_{*}$.  The "hot" debris disks in the range 600K-200K include a significant amount of emission towards the red end of the S9W filter (10-12 microns.)  The traditional definition of a debris disk is $f<0.01$, so for an M4 (3200K) dwarf the S9W detection limit lies above the debris disk regime. For hotter stars which make up the bulk of the sample, we are sensitive to massive debris disks that lie close to the star, just inwards from the "habitable zone."  Planetary collisions or other transient events might also produce large $f$. Although the L18W filter appears more sensitive, that is somewhat misleading since only a few objects are detected at L18W. 

\begin{figure}
\epsscale{0.7}
\plotone{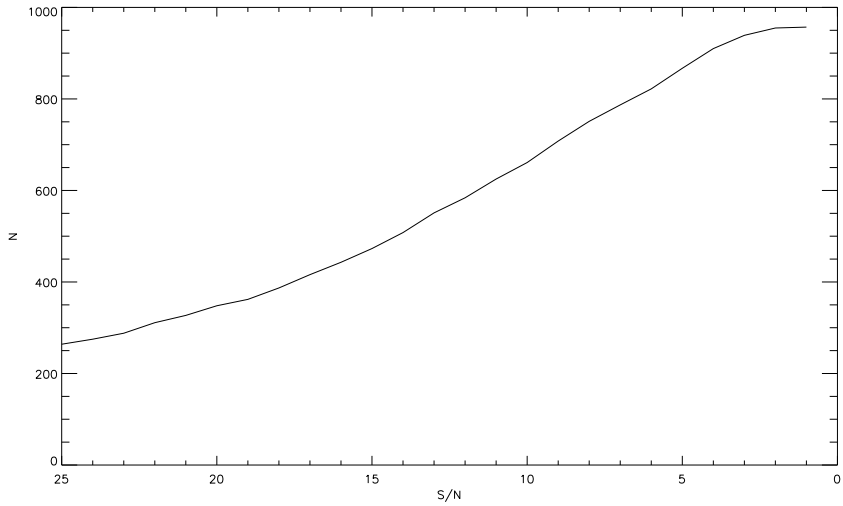}
\caption{The number of stars with at least a given signal-to-noise in the S9W band.  
\label{fig-cdf}}
\end{figure}

As noted previously, the AKARI-2MASS data shows considerable scatter, more than one would expect from the reported uncertainties. While metallicity variations and other intrinsic scatter may be contributing, it seems likely that the errors are simply underestimated. Despite the scatter,we find no good debris disk candidates in Tables~2 and~\ref{tab-reid} until near, or below, the S9W 0.4 magnitude limit plotted in Figures~\ref{fig-fdet}. None lie above the L18W 0.5 magnitude limits ~\ref{fig-fdet18}.  Neither magnetically active (dMe) nor inactive (dM) stars have such large excesses.  Figure~\ref{fig-cdf} shows the distribution of AKARI signal-to-noise measurements for the Table~2 sample. (We also caution again that many excesses may be found by naively matching AKARI and 2MASS, but these are due to double 2MASS sources matched to single AKARI sources.)  Overall 191 detected stars in Table~2 have H$\alpha$ emission and 748 do not.  Although dominated by old and single stars, the sample necessarily includes young field stars and binary systems with separations less than $\sim 3$ arcseconds, or $\sim 50$ AU.  

For comparison, \citet{2009A&A...506.1455L}'s analysis of submillimeter data implies that $5.3^{+10.5}_{?5.0}\%$of "young" ($<200$ Myr) have cold debris disks and  $<10\%$ of "old" ($\gtrsim 1$ Gyr) M dwarfs have cold debris disks. \citet{2007ApJ...667..527G} note that no examples of warm (24,70 or 160 micron) debris disks around "mature" M dwarfs are known and place a upper limit of 14\% n their frequency from their twenty stars. This sample of M dwarfs within 5 parsecs is dominated  by older stars. However, \citet{2009AAS...21430105P} report the discovery of three M dwarf debris disks with Spittzer/MIPS, but the specific objects have not yet been published, so the fraction is unknown. 

The detection limits seen in Figure~\ref{fig-fdet} are rather large; nevertheless, the search for bright disks is valuable. \citet{2008ARA&A..46..339W} describes cases where the presence of hot dust around more massive stars is difficult to understand. For M dwarfs, \citet{2005ApJ...631.1161P} noted that "the characteristic grain removal timescales from [Poynting-Robertson] drag are significantly longer than for earlier type dwarfs due to the relatively lower luminosities in late-type dwarfs." They also showed dust grains cannot be blown out by M dwarf radiation pressure. These considerations suggest that hot dust grains could be common around M dwarfs. Since, however, they did not find any excesses in their young M dwarf sample at $11.7\mu$mm, they proposed that stellar wing drag might be responsible for removing grains. They predict  $f \approx 10^{-6}$, well below our reliable, but conservative, detection limits above, if this model is correct. The present AKARI-based results so far seem support that model. However, as seen in Figure~\ref{fig-photo}, if one pushes to lower detection thresholds in S9W there are candidate excesses.

The best candidates for hot debris disks are listed in Table~4.  LHS 4058, Gl 121.1, LHS 3799, Gl 821, and Gl 828.2  all have high signal-noise-measurements in AKARI and are $\sim 30\%-40\%$ brighter than expected. {\it These excesses are more than ten times the error due to AKARI alone.} On the other hand, the observed scatter is larger, and given the large sample we expect some outliers, so independent confirmation is required. None of these are detected in the L18W band but with $m_{S9W} > 6.3$, detections are not expected.  \citet{2008MNRAS.389..585C} provide a detailed analysis of the effective temperatures and metallicities of a large sample of M dwarfs, including Gl 828.2 and Gl 821. For those two stars, their luminosities and temperatures in Table~4 and use them to calculate $f$. For the other three stars, we estimate temperatures using Equation~\ref{eqn-temp} and  \citet{2008MNRAS.389..585C}'s equation $M_{bol} = 2.07 + 1.08 \times M_K$. For LHS 4058, the given distance is the PMSU spectroscopic estimate; the other four have Hipparcos parallaxes. For ages, an upper limit (for the H$\alpha$ emission star) or lower limit (for the non-emission stars) are given using the prescription of \citet{2008AJ....135..785W}. If all five truly have debris disks, then the overall M dwarf debris disk fraction would be $\sim 1\%$. \citet{2008MNRAS.389..585C} claim that Gl 828.2 and Gl 821  
are metal-poor ($[m/H] = -0.52 \pm 0.2$ and $-0.65 \pm 0.2$, respectively) which would make a massive debris disk rather surprising. \citet{2008MNRAS.389..585C} discuss the challenge of estimating metallicities from broad-band colors in detail. Even if all five have hot debris disks, we can still say with 95\% confidence, that frequency of hot, bright debris disks is $<2\%$.  

A number of additional stars deserve discussion. The stars Gl 526, Gl 799 (AT Mic), Gl 803 (AU Mic), Gl 842.2, and Gl 907.1 have confirmed or suspected debris disks at longer wavelengths but definitely have no excess at S9W. \citet{2009ApJ...698.1068P} find AT Mic (Gl 799A), a $\beta$ Pic moving group member,  is 15\% brighter at 24 microns than expected. This corresponds to only 0.15 magnitudes, difficult for us to recognize given the apparent scatter. We find $K_s - m_{L18W} = 0.86$, implying an excess of 0.11 magnitude compared to the NextGen models, consistent with the better Spitzer data. They also find no 24 micron excess for AU Mic, consistent with our detection at L18W ($K_s - m_{L18W} = 0.29$).  Ten of our detected systems --- Gl 179 \citep{2010arXiv1003.3488H}, Gl 317 \citep{2007ApJ...670..833J}, Gl 436 \citep{2004ApJ...617..580B}, Gl 581 \citep{2005A&A...443L..15B}, Gl 649 \citep{2010PASP..122..149J}, Gl 674 \citep{2007A&A...474..293B}, 832 \citep{2009ApJ...690..743B}, Gl 849 \citep{2006PASP..118.1685B}, Gl 876 \citep{2005ApJ...634..625R} and GJ 1148 \citep{2010ApJ...715..271H} --- have known planets: none are even 0.1 magnitude brighter than expected, ruling out hot debris disks.  
 
\section{Conclusion}

The better resolution and sensitivity of the AKARI sky survey relative to IRAS allows a sample of nearly a thousand late-type K5-M6 dwarfs detected in a broad filter centered at 9 microns to be identified, with fifty detected at 18 microns. The observed brightnesses of the vast majority are consistent with the model photosphere predictions. Five stars with apparent excesses of $\sim 30$\% in the S9W micron filter are identified. Given the likelihood of a few extreme outliers in a large sample, they require further follow-up to be confirmed as hot debris disk systems.  In any case, hot debris disks are rare. In the 18 micron filter, a 10\% excess is observed for the star AT Mic, which is consistent with a previously reported excess at 24 microns.  

\acknowledgments

I thank George Helou and the Spitzer Science Center for a very educational visit during my sabbatical. I would also like to thank Kelle Cruz and the writers of the \url{AstroBetter.com} for helpful tips and tricks.  
I thank the Annie Jump Cannon Fund at the University of Delaware for support.

This research is based on observations with AKARI, a JAXA project with the participation of ESA. This publication makes use of data products from the Two Micron All Sky Survey, which is a joint project of the University of Massachusetts and the Infrared Processing and Analysis Center/California Institute of Technology, funded by the National Aeronautics and Space Administration and the National Science Foundation. This research has made use of the NASA/ IPAC Infrared Science Archive, which is operated by the Jet Propulsion Laboratory, California Institute of Technology, under contract with NASA.This research has made use of the VizieR catalogue access tool, CDS, Strasbourg, France. This research has made use of the SIMBAD database, operated at CDS, Strasbourg, France.

\bibliographystyle{apj}
\bibliography{../astrobib}

\begin{thebibliography}{47}
\expandafter\ifx\csname natexlab\endcsname\relax\def\natexlab#1{#1}\fi

\bibitem[{{Bailey} {et~al.}(2009){Bailey}, {Butler}, {Tinney}, {Jones},
  {O'Toole}, {Carter}, \& {Marcy}}]{2009ApJ...690..743B}
{Bailey}, J., {Butler}, R.~P., {Tinney}, C.~G., {Jones}, H.~R.~A., {O'Toole},
  S., {Carter}, B.~D., \& {Marcy}, G.~W. 2009, \apj, 690, 743

\bibitem[{{Beichman}(1987)}]{1987ARA&A..25..521B}
{Beichman}, C.~A. 1987, \araa, 25, 521

\bibitem[{{Bonfils} {et~al.}(2005){Bonfils}, {Forveille}, {Delfosse}, {Udry},
  {Mayor}, {Perrier}, {Bouchy}, {Pepe}, {Queloz}, \&
  {Bertaux}}]{2005A&A...443L..15B}
{Bonfils}, X., {et~al.} 2005, \aap, 443, L15

\bibitem[{{Bonfils} {et~al.}(2007){Bonfils}, {Mayor}, {Delfosse}, {Forveille},
  {Gillon}, {Perrier}, {Udry}, {Bouchy}, {Lovis}, {Pepe}, {Queloz}, {Santos},
  \& {Bertaux}}]{2007A&A...474..293B}
---. 2007, \aap, 474, 293

\bibitem[{{Butler} {et~al.}(2006){Butler}, {Johnson}, {Marcy}, {Wright},
  {Vogt}, \& {Fischer}}]{2006PASP..118.1685B}
{Butler}, R.~P., {Johnson}, J.~A., {Marcy}, G.~W., {Wright}, J.~T., {Vogt},
  S.~S., \& {Fischer}, D.~A. 2006, \pasp, 118, 1685

\bibitem[{{Butler} {et~al.}(2004){Butler}, {Vogt}, {Marcy}, {Fischer},
  {Wright}, {Henry}, {Laughlin}, \& {Lissauer}}]{2004ApJ...617..580B}
{Butler}, R.~P., {Vogt}, S.~S., {Marcy}, G.~W., {Fischer}, D.~A., {Wright},
  J.~T., {Henry}, G.~W., {Laughlin}, G., \& {Lissauer}, J.~J. 2004, \apj, 617,
  580

\bibitem[{{Casagrande} {et~al.}(2008){Casagrande}, {Flynn}, \&
  {Bessell}}]{2008MNRAS.389..585C}
{Casagrande}, L., {Flynn}, C., \& {Bessell}, M. 2008, \mnras, 389, 585

\bibitem[{{Engelbracht} {et~al.}(2007){Engelbracht}, {Blaylock}, {Su}, {Rho},
  {Rieke}, {Muzerolle}, {Padgett}, {Hines}, {Gordon}, {Fadda},
  {Noriega-Crespo}, {Kelly}, {Latter}, {Hinz}, {Misselt}, {Morrison},
  {Stansberry}, {Shupe}, {Stolovy}, {Wheaton}, {Young}, {Neugebauer},
  {Wachter}, {P{\'e}rez-Gonz{\'a}lez}, {Frayer}, \&
  {Marleau}}]{2007PASP..119..994E}
{Engelbracht}, C.~W., {et~al.} 2007, \pasp, 119, 994

\bibitem[{{Forbrich} {et~al.}(2008){Forbrich}, {Lada}, {Muench}, \&
  {Teixeira}}]{2008ApJ...687.1107F}
{Forbrich}, J., {Lada}, C.~J., {Muench}, A.~A., \& {Teixeira}, P.~S. 2008,
  \apj, 687, 1107

\bibitem[{{Fujiwara} {et~al.}(2010){Fujiwara}, {Onaka}, {Ishihara},
  {Yamashita}, {Fukagawa}, {Nakagawa}, {Kataza}, {Ootsubo}, \&
  {Murakami}}]{2010ApJ...714L.152F}
{Fujiwara}, H., {et~al.} 2010, \apjl, 714, L152

\bibitem[{{Furlan} {et~al.}(2007){Furlan}, {Sargent}, {Calvet}, {Forrest},
  {D'Alessio}, {Hartmann}, {Watson}, {Green}, {Najita}, \&
  {Chen}}]{2007ApJ...664.1176F}
{Furlan}, E., {et~al.} 2007, \apj, 664, 1176

\bibitem[{{Gautier} {et~al.}(2008){Gautier}, {Rebull}, {Stapelfeldt}, \&
  {Mainzer}}]{2008ApJ...683..813G}
{Gautier}, III, T.~N., {Rebull}, L.~M., {Stapelfeldt}, K.~R., \& {Mainzer}, A.
  2008, \apj, 683, 813

\bibitem[{{Gautier} {et~al.}(2007){Gautier}, {Rieke}, {Stansberry}, {Bryden},
  {Stapelfeldt}, {Werner}, {Beichman}, {Chen}, {Su}, {Trilling}, {Patten}, \&
  {Roellig}}]{2007ApJ...667..527G}
{Gautier}, III, T.~N., {et~al.} 2007, \apj, 667, 527

\bibitem[{{Gliese} \& {Jahrei{\ss}}(1991)}]{pcns3}
{Gliese}, W., \& {Jahrei{\ss}}, H. 1991, {Preliminary Version of the Third
  Catalogue of Nearby Stars}, Tech. rep., Astronomisches Rechen-Institut

\bibitem[{{Haghighipour} {et~al.}(2010){Haghighipour}, {Vogt}, {Butler},
  {Rivera}, {Laughlin}, {Meschiari}, \& {Henry}}]{2010ApJ...715..271H}
{Haghighipour}, N., {Vogt}, S.~S., {Butler}, R.~P., {Rivera}, E.~J.,
  {Laughlin}, G., {Meschiari}, S., \& {Henry}, G.~W. 2010, \apj, 715, 271

\bibitem[{{Hauschildt} {et~al.}(1999){Hauschildt}, {Allard}, \&
  {Baron}}]{1999ApJ...512..377H}
{Hauschildt}, P.~H., {Allard}, F., \& {Baron}, E. 1999, \apj, 512, 377

\bibitem[{{Hawley} {et~al.}(1996){Hawley}, {Gizis}, \&
  {Reid}}]{1996AJ....112.2799H}
{Hawley}, S.~L., {Gizis}, J.~E., \& {Reid}, I.~N. 1996, \aj, 112, 2799

\bibitem[{{Howard} {et~al.}(2010){Howard}, {Johnson}, {Marcy}, {Fischer},
  {Wright}, {Bernat}, {Henry}, {Peek}, {Isaacson}, {Apps}, {Endl}, {Cochran},
  {Valenti}, {Anderson}, \& {Piskunov}}]{2010arXiv1003.3488H}
{Howard}, A.~W., {et~al.} 2010, ArXiv e-prints

\bibitem[{{Ishihara} {et~al.}(2010){Ishihara}, {Onaka}, {Kataza}, {Salama},
  {Alfageme}, {Cassatella}, {Cox}, {Garc{\'{\i}}a-Lario}, {Stephenson},
  {Cohen}, {Fujishiro}, {Fujiwara}, {Hasegawa}, {Ita}, {Kim}, {Matsuhara},
  {Murakami}, {M{\"u}ller}, {Nakagawa}, {Ohyama}, {Oyabu}, {Pyo}, {Sakon},
  {Shibai}, {Takita}, {Tanab{\'e}}, {Uemizu}, {Ueno}, {Usui}, {Wada},
  {Watarai}, {Yamamura}, \& {Yamauchi}}]{akari-psc}
{Ishihara}, D., {et~al.} 2010, \aap, 514, A1+

\bibitem[{{Ita} {et~al.}(2010){Ita}, {Matsuura}, {Ishihara}, {Oyabu}, {Takita},
  {Kataza}, {Yamamura}, {Matsunaga}, {Tanab{\'e}}, {Nakada}, {Fujiwara},
  {Wada}, {Onaka}, \& {Matsuhara}}]{akari-nearby}
{Ita}, Y., {et~al.} 2010, \aap, 514, A2+

\bibitem[{{Johnson} {et~al.}(2007){Johnson}, {Butler}, {Marcy}, {Fischer},
  {Vogt}, {Wright}, \& {Peek}}]{2007ApJ...670..833J}
{Johnson}, J.~A., {Butler}, R.~P., {Marcy}, G.~W., {Fischer}, D.~A., {Vogt},
  S.~S., {Wright}, J.~T., \& {Peek}, K.~M.~G. 2007, \apj, 670, 833

\bibitem[{{Johnson} {et~al.}(2010){Johnson}, {Howard}, {Marcy}, {Bowler},
  {Henry}, {Fischer}, {Apps}, {Isaacson}, \& {Wright}}]{2010PASP..122..149J}
{Johnson}, J.~A., {et~al.} 2010, \pasp, 122, 149

\bibitem[{{Kalas} {et~al.}(2004){Kalas}, {Liu}, \&
  {Matthews}}]{2004Sci...303.1990K}
{Kalas}, P., {Liu}, M.~C., \& {Matthews}, B.~C. 2004, Science, 303, 1990

\bibitem[{Kataza(2010)}]{akari-psc-note}
Kataza, H., e.~a. 2010, AKARI-IRC Point Source Catalogue Release note Version
  1.0, Tech. rep., ISAS/JAXA

\bibitem[{{Lestrade} {et~al.}(2006){Lestrade}, {Wyatt}, {Bertoldi}, {Dent}, \&
  {Menten}}]{2006A&A...460..733L}
{Lestrade}, J., {Wyatt}, M.~C., {Bertoldi}, F., {Dent}, W.~R.~F., \& {Menten},
  K.~M. 2006, \aap, 460, 733

\bibitem[{{Lestrade} {et~al.}(2009){Lestrade}, {Wyatt}, {Bertoldi}, {Menten},
  \& {Labaigt}}]{2009A&A...506.1455L}
{Lestrade}, J., {Wyatt}, M.~C., {Bertoldi}, F., {Menten}, K.~M., \& {Labaigt},
  G. 2009, \aap, 506, 1455

\bibitem[{{Liu} {et~al.}(2004){Liu}, {Matthews}, {Williams}, \&
  {Kalas}}]{2004ApJ...608..526L}
{Liu}, M.~C., {Matthews}, B.~C., {Williams}, J.~P., \& {Kalas}, P.~G. 2004,
  \apj, 608, 526

\bibitem[{{Mainzer} {et~al.}(2007){Mainzer}, {Roellig}, {Saumon}, {Marley},
  {Cushing}, {Sloan}, {Kirkpatrick}, {Leggett}, \&
  {Wilson}}]{2007ApJ...662.1245M}
{Mainzer}, A.~K., {et~al.} 2007, \apj, 662, 1245

\bibitem[{{Mamajek} {et~al.}(1999){Mamajek}, {Lawson}, \&
  {Feigelson}}]{1999ApJ...516L..77M}
{Mamajek}, E.~E., {Lawson}, W.~A., \& {Feigelson}, E.~D. 1999, \apjl, 516, L77

\bibitem[{{Megeath} {et~al.}(2005){Megeath}, {Hartmann}, {Luhman}, \&
  {Fazio}}]{2005ApJ...634L.113M}
{Megeath}, S.~T., {Hartmann}, L., {Luhman}, K.~L., \& {Fazio}, G.~G. 2005,
  \apjl, 634, L113

\bibitem[{{Murakami} {et~al.}(2007){Murakami}, {Baba}, {Barthel}, {Clements},
  {Cohen}, {Doi}, {Enya}, {Figueredo}, {Fujishiro}, {Fujiwara}, {Fujiwara},
  {Garcia-Lario}, {Goto}, {Hasegawa}, {Hibi}, {Hirao}, {Hiromoto}, {Hong},
  {Imai}, {Ishigaki}, {Ishiguro}, {Ishihara}, {Ita}, {Jeong}, {Jeong},
  {Kaneda}, {Kataza}, {Kawada}, {Kawai}, {Kawamura}, {Kessler}, {Kester},
  {Kii}, {Kim}, {Kim}, {Kobayashi}, {Koo}, {Kwon}, {Lee}, {Lorente}, {Makiuti},
  {Matsuhara}, {Matsumoto}, {Matsuo}, {Matsuura}, {M{\"u}ller}, {Murakami},
  {Nagata}, {Nakagawa}, {Naoi}, {Narita}, {Noda}, {Oh}, {Ohnishi}, {Ohyama},
  {Okada}, {Okuda}, {Oliver}, {Onaka}, {Ootsubo}, {Oyabu}, {Pak}, {Park},
  {Pearson}, {Rowan-Robinson}, {Saito}, {Sakon}, {Salama}, {Sato}, {Savage},
  {Serjeant}, {Shibai}, {Shirahata}, {Sohn}, {Suzuki}, {Takagi}, {Takahashi},
  {Tanab{\'e}}, {Takeuchi}, {Takita}, {Thomson}, {Uemizu}, {Ueno}, {Usui},
  {Verdugo}, {Wada}, {Wang}, {Watabe}, {Watarai}, {White}, {Yamamura},
  {Yamauchi}, \& {Yasuda}}]{akari}
{Murakami}, H., {et~al.} 2007, \pasj, 59, 369

\bibitem[{{Onaka} {et~al.}(2007){Onaka}, {Matsuhara}, {Wada}, {Fujishiro},
  {Fujiwara}, {Ishigaki}, {Ishihara}, {Ita}, {Kataza}, {Kim}, {Matsumoto},
  {Murakami}, {Ohyama}, {Oyabu}, {Sakon}, {Tanab{\'e}}, {Takagi}, {Uemizu},
  {Ueno}, {Usui}, {Watarai}, {Cohen}, {Enya}, {Ootsubo}, {Pearson}, {Takeyama},
  {Yamamuro}, \& {Ikeda}}]{akari-irc}
{Onaka}, T., {et~al.} 2007, \pasj, 59, 401

\bibitem[{{Perryman} {et~al.}(1997){Perryman}, {Lindegren}, {Kovalevsky},
  {Hoeg}, {Bastian}, {Bernacca}, {Cr{\'e}z{\'e}}, {Donati}, {Grenon}, {van
  Leeuwen}, {van der Marel}, {Mignard}, {Murray}, {Le Poole}, {Schrijver},
  {Turon}, {Arenou}, {Froeschl{\'e}}, \& {Petersen}}]{hipparcos}
{Perryman}, M.~A.~C., {et~al.} 1997, \aap, 323, L49

\bibitem[{{Plavchan} {et~al.}(2009{\natexlab{a}}){Plavchan}, {Bryden},
  {Werner}, {Rieke}, {Stapelfeldt}, {Lowrance}, {Lestrade}, \&
  {Tanner}}]{2009AAS...21430105P}
{Plavchan}, P., {Bryden}, G., {Werner}, M., {Rieke}, G., {Stapelfeldt}, K.,
  {Lowrance}, P., {Lestrade}, J., \& {Tanner}, A. 2009{\natexlab{a}}, in
  American Astronomical Society Meeting Abstracts, Vol. 214, American
  Astronomical Society Meeting Abstracts, 301.05--+

\bibitem[{{Plavchan} {et~al.}(2005){Plavchan}, {Jura}, \&
  {Lipscy}}]{2005ApJ...631.1161P}
{Plavchan}, P., {Jura}, M., \& {Lipscy}, S.~J. 2005, \apj, 631, 1161

\bibitem[{{Plavchan} {et~al.}(2009{\natexlab{b}}){Plavchan}, {Werner}, {Chen},
  {Stapelfeldt}, {Su}, {Stauffer}, \& {Song}}]{2009ApJ...698.1068P}
{Plavchan}, P., {Werner}, M.~W., {Chen}, C.~H., {Stapelfeldt}, K.~R., {Su},
  K.~Y.~L., {Stauffer}, J.~R., \& {Song}, I. 2009{\natexlab{b}}, \apj, 698,
  1068

\bibitem[{{Reid} \& {Hawley}(2005)}]{2005-book-reid}
{Reid}, I.~N., \& {Hawley}, S.~L. 2005, {New light on dark stars : red dwarfs,
  low-mass stars, brown dwarfs} (Springer)

\bibitem[{{Reid} {et~al.}(1995){Reid}, {Hawley}, \&
  {Gizis}}]{1995AJ....110.1838R}
{Reid}, I.~N., {Hawley}, S.~L., \& {Gizis}, J.~E. 1995, \aj, 110, 1838

\bibitem[{{Reid} {et~al.}(2004){Reid}, {Cruz}, {Allen}, {Mungall}, {Kilkenny},
  {Liebert}, {Hawley}, {Fraser}, {Covey}, {Lowrance}, {Kirkpatrick}, \&
  {Burgasser}}]{2004AJ....128..463R}
{Reid}, I.~N., {et~al.} 2004, \aj, 128, 463

\bibitem[{{Rhee} {et~al.}(2007){Rhee}, {Song}, {Zuckerman}, \&
  {McElwain}}]{2007ApJ...660.1556R}
{Rhee}, J.~H., {Song}, I., {Zuckerman}, B., \& {McElwain}, M. 2007, \apj, 660,
  1556

\bibitem[{{Riaz} {et~al.}(2006{\natexlab{a}}){Riaz}, {Gizis}, \&
  {Harvin}}]{2006AJ....132..866R}
{Riaz}, B., {Gizis}, J.~E., \& {Harvin}, J. 2006{\natexlab{a}}, \aj, 132, 866

\bibitem[{{Riaz} {et~al.}(2006{\natexlab{b}}){Riaz}, {Mullan}, \&
  {Gizis}}]{2006ApJ...650.1133R}
{Riaz}, B., {Mullan}, D.~J., \& {Gizis}, J.~E. 2006{\natexlab{b}}, \apj, 650,
  1133

\bibitem[{{Rivera} {et~al.}(2005){Rivera}, {Lissauer}, {Butler}, {Marcy},
  {Vogt}, {Fischer}, {Brown}, {Laughlin}, \& {Henry}}]{2005ApJ...634..625R}
{Rivera}, E.~J., {et~al.} 2005, \apj, 634, 625

\bibitem[{{Skrutskie} {et~al.}(2006){Skrutskie}, {Cutri}, {Stiening},
  {Weinberg}, {Schneider}, {Carpenter}, {Beichman}, {Capps}, {Chester},
  {Elias}, {Huchra}, {Liebert}, {Lonsdale}, {Monet}, {Price}, {Seitzer},
  {Jarrett}, {Kirkpatrick}, {Gizis}, {Howard}, {Evans}, {Fowler}, {Fullmer},
  {Hurt}, {Light}, {Kopan}, {Marsh}, {McCallon}, {Tam}, {Van Dyk}, \&
  {Wheelock}}]{2mass}
{Skrutskie}, M.~F., {et~al.} 2006, \aj, 131, 1163

\bibitem[{{Song} {et~al.}(2002){Song}, {Weinberger}, {Becklin}, {Zuckerman}, \&
  {Chen}}]{2002AJ....124..514S}
{Song}, I., {Weinberger}, A.~J., {Becklin}, E.~E., {Zuckerman}, B., \& {Chen},
  C. 2002, \aj, 124, 514

\bibitem[{{West} {et~al.}(2008){West}, {Hawley}, {Bochanski}, {Covey}, {Reid},
  {Dhital}, {Hilton}, \& {Masuda}}]{2008AJ....135..785W}
{West}, A.~A., {Hawley}, S.~L., {Bochanski}, J.~J., {Covey}, K.~R., {Reid},
  I.~N., {Dhital}, S., {Hilton}, E.~J., \& {Masuda}, M. 2008, \aj, 135, 785

\bibitem[{{Wyatt}(2008)}]{2008ARA&A..46..339W}
{Wyatt}, M.~C. 2008, \araa, 46, 339

\end{thebibliography}



\clearpage
\begin{deluxetable}{llrrrrr}
\tablewidth{0pc}
\tablenum{1}\label{tab-astars}

\tablecaption{A Stars}
\tablehead{
\colhead{HD Name} &
\colhead{AKARI Name} &
\colhead{J} &
\colhead{H} &
\colhead{K$_s$} & 
\colhead{m$_{S9W}$} & 
\colhead{m$_{L18W}$} 
}
\startdata
HD 319 & 0007469-223031 & 5.601 & 5.509 & 5.479 & 5.45 & \nodata \\
HD 11413 & 0150543-501221 & 5.544 & 5.438 & 5.422 & 5.38 & \nodata \\ 
HD 21981 & 0330370-472230 & 5.651 & 5.621 & 5.526 & 5.53 & \nodata \\ 
HD 22686 & 0338551+024547 & 7.159 & 7.199 & 7.132 & 6.88 & \nodata \\ 
HD 45557 & 0624138-601652 & 5.755 & 5.803 & 5.754 & 5.69 & \nodata \\ 
HD 46190 & 0627486-620859 & 6.411 & 6.391 & 6.363 & 6.32 & \nodata \\ 
HD 58142 & 0726427+491240 & 4.695 & 4.693 & 4.572 & 4.60 & 4.59 \\ 
HD 73666 & 0840114+195815 & 6.498 & 6.547 & 6.532 & 6.37 & \nodata \\ 
HD 91375 & 1030201-715934 & 4.747 & 4.715 & 4.565 & 4.59 & 4.58 \\ 
HD 95418 & 1101506+562257 & 2.269 & 2.359 & 2.285 & 2.38 & 2.22 \\ 
HD 128998 & 1438152+540123 & 5.756 & 5.801 & 5.756 & 5.78 & \nodata \\ 
\enddata
\end{deluxetable}

\begin{deluxetable}{rllrrrrrrrrrrrr}
\tabletypesize{\footnotesize}
\rotate
\tablewidth{10in}
\tablenum{2}\label{tab-main}
\tablecaption{Nearby M Dwarfs}
\tablehead{
\colhead{[RHG95]} &
\colhead{Name} &
\colhead{AKARI} &
\colhead{J} &
\colhead{H} &
\colhead{K$_{s}$} & 
\colhead{m$_{S9W}$} & 
\colhead{m$_{L18W}$} & 
\colhead{SN$_{S9W}$} & 
\colhead{SN$_{L18W}$} & 
\colhead{Sp.T.} &
\colhead{TiO5} &
\colhead{H$\alpha$} &
\colhead{$\Delta_{9}$} &
\colhead{$N_\sigma$} 
}
\startdata
 3796 & GJ 1293       & 0001259-165655 & 8.02 & 7.41 & 7.22 & 6.88 & \nodata & 9.8 & \nodata & M0 & 0.74 & 0.00 & 0.06 & 0.6 \\ 
 10 & BPM16002      & 0005249-500252 & 8.55 & 7.91 & 7.63 & 7.93 & \nodata & 2.4 & \nodata & M1.5 & 0.62 & 0.00 & -0.65 & 1.7 \\ 
 6 & Gl 1          & 0005277-372142 & 5.33 & 4.83 & 4.52 & 4.21 & 3.98 & 84.6 & 9.9 & M1.5 & 0.61 & 0.00 & -0.04 & 3.1 \\ 
 15 & LHS1019       & 0006192-655027 & 8.48 & 7.84 & 7.63 & 7.44 & \nodata & 6.1 & \nodata & M2.5 & 0.53 & 0.00 & -0.21 & 1.3 \\ 
 19 & GJ 1002       & 0006428-073230 & 8.32 & 7.79 & 7.44 & 6.49 & \nodata & 7.9 & \nodata & M5.5 & 0.24 & 0.00 & 0.37 & 2.9 \\ 
 24 & LTT 57        & 0008172-570553 & 8.21 & 7.63 & 7.39 & 6.93 & \nodata & 6.3 & \nodata & M2.0 & 0.49 & 0.00 & 0.04 & 0.2 \\ 
 23 & V 351         & 0008272+172526 & 7.81 & 7.17 & 6.98 & 6.72 & \nodata & 11.1 & \nodata & M0 & 0.77 & 0.00 & 0.00 & 0.0 \\ 
 30 & G 031-030     & 0011047-054702 & 7.98 & 7.31 & 7.12 & 6.55 & \nodata & 5.5 & \nodata & K7 & 0.79 & 0.00 & 0.32 & 1.8 \\
\enddata
\end{deluxetable}

\begin{deluxetable}{llrrrrr}
\tablewidth{0pc}
\tablenum{3}\label{tab-reid}

\tablecaption{Additional Nearby Dwarfs}
\tablehead{
\colhead{Name} &
\colhead{AKARI} &
\colhead{J} &
\colhead{H} &
\colhead{K$_{s}$} & 
\colhead{m$_{S9W}$} & 
\colhead{SN$_{S9W}$}  
}
\startdata
BD-14 249    & 0117159-131547 & 7.78 & 7.18 & 6.92 & 6.65 & 22.3 \\ 
LP 130- 37   & 0428058-620925 & 8.93 & 8.36 & 8.01 & 7.54 & 12.8  \\ 
L 381-4      & 0636181-400022 & 7.64 & 7.07 & 6.79 & 5.99 & 52.8  \\ 
LP 32-  9    & 0633427-753745 & 6.74 & 6.14 & 5.86 & 5.48 & 37.7  \\ 
BD-2 2198    & 0736071-030640 & 6.81 & 6.14 & 5.94 & 5.75 & 12.2  \\ 
LP 784- 12   & 0802031-171018 & 8.19 & 7.61 & 7.36 & 7.22 & 33.3 \\ 
LP 843- 17   & 0815401-260037 & 7.62 & 7.06 & 6.86 & 6.60 & 4.9  \\ 
CD-31 13358  & 1655380-320406 & 6.89 & 6.23 & 6.03 & 5.87 & 13.0 \\
\enddata
\end{deluxetable}

\begin{deluxetable}{lrrrrr}
\tablewidth{0pc}
\tablenum{4}\label{tab-can}
\tablecaption{Candidates}
\tablehead{
\colhead{Name} &
\colhead{d (pc)} &
\colhead{$L/L_\odot$} &
\colhead{$T_{eff}$ (K)} &
\colhead{f (400K)} &
\colhead{Age (Gyr)} }
\startdata
Gl 121.1  & 20.9 & 0.038 & 3567 & 0.008 & $>1.2$ \\
Gl 828.2  &  16.2 & 0.026 & 3585 & 0.006 & $>0.4$ \\
Gl 821   & 12.1 & 0.019 &  3567 & 0.006 & $>0.4$ \\
LHS 3799 & 7.5 & 0.0042 & 3100 & 0.010 &  $<5.8$ \\
LHS 4058 & 12.3 & 0.008 & 3350 & 0.010 & $>2.0$ \\
\enddata
\end{deluxetable}

\end{document}